\documentclass[preprint,prd,nofootinbib,tightenlines,amsmath]{revtex4}
\usepackage{graphicx}% Include figure files
\usepackage{bm}% bold math

\begin{document}
\baselineskip=15pt \parskip=4pt

\vspace*{3em}

\title{Anomalous $\bm{CP}$-Violation in $\bm{B_s}$-$\bm{\bar{B}_s}$ Mixing \\
Due to a Light Spin-One Particle}

\author{Sechul Oh$^{1,3}$}
\author{Jusak Tandean$^{2,3}$}

\affiliation{
$^1$Institute of Physics, Academia Sinica, \\ Taipei 115, Taiwan  \vspace*{1ex} \\
$^2$Center for Mathematics and Theoretical Physics and Department of Physics, \\
National Central University, Chungli 320, Taiwan \vspace*{1ex} \\
$^3$Department of Physics and Center for Theoretical Sciences, \\
National Taiwan University, Taipei 106, Taiwan \\
$\vphantom{\bigg|_{\bigg|}^|}$}

%\date{\today}

\begin{abstract}
The recent measurement of the like-sign dimuon charge asymmetry in semileptonic $b$-hadron
decays by the D0 Collaboration is about three sigmas away from the standard-model prediction,
hinting at the presence of $CP$-violating new physics in the mixing of $B_s$ mesons.
We consider the possibility that this anomalous result arises from the contribution of
a light spin-1 particle.
Taking into account various experimental constraints, we find that the effect of such
a particle with mass below the $b$-quark mass can yield a~prediction consistent with
the anomalous D0 measurement within its one-sigma range.
\end{abstract}

\maketitle

\section{Introduction}

The D0 Collaboration has recently reported a new measurement of the like-sign
dimuon charge asymmetry in semileptonic $b$-hadron
decays,~\,$A^b_{\rm sl}=[-9.57\pm2.51\,\rm(stat)\pm1.46\,(sys)]\times10^{-3}$\,~\cite{Abazov:2010hv}.
It disagrees with the standard model (SM) prediction
\,$A^{b,\rm SM}_{\rm sl}=\bigl(-2.3^{+0.5}_{-0.6}\bigr)\times 10^{-4}$\,~\cite{Beneke:1998sy,Lenz:2006hd}
by 3.2~standard deviations, thereby providing evidence
for anomalous $CP$-violation in the mixing of neutral \mbox{$B$-mesons}.
This observable is related to the charge asymmetry $a^s_{\rm sl}$ for ``wrong-charge''
semileptonic $B_s$ decay induced by oscillations.
The above values of $A^b_{\rm sl}$ thus translate
into~\cite{Abazov:2010hv,Lenz:2006hd}
\begin{eqnarray}
a^{s,\rm exp}_{\rm sl} \,&=&\, -(14.6 \pm 7.5) \times 10^{-3} ~, \label{asslx} \\
a^{s,\rm SM}_{\rm sl} \,&=&\, (2.1 \pm 0.6) \times 10^{-5} ~. \label{assl_sm}
\end{eqnarray}
Although not yet conclusive, this sizable discrepancy between experiment and theory suggests
that new physics beyond the SM may be responsible for it.
Consequently, it has attracted a great deal of attention in the
literature~\cite{D0_anomaly,Dobrescu:2010rh,Kubo:2010mh}.

In addition to $a^s_{\rm sl}$, the observables of interest in this case are
the mass and width differences $\Delta M_s$ and $\Delta\Gamma_s$, respectively,
between the heavy and light mass-eigenstates in the $B_s$-$\bar B_s$ system.
Their experimental values are~\cite{pdg}
\begin{eqnarray} \label{DeltaMs_data}
\Delta M_s^{\rm exp} \,\,=\,\, 17.77\pm 0.12 ~{\rm ps}^{-1} ~, \hspace{5ex}
\Delta\Gamma_s^{\rm exp} \,\,=\,\, 0.062_{-0.037}^{+0.034} ~{\rm ps}^{-1} ~.
\end{eqnarray}
These three observables are related to the off-diagonal elements $M_s^{12}$ and $\Gamma_s^{12}$
of the mass and decay matrices, respectively, which characterize $B_s$-$\bar B_s$ mixing.
The relationship is described by~\cite{Branco:1999fs}
\begin{eqnarray} \label{DeltaMs}
&& \hspace*{-5ex} \bigl(\Delta M_s^{}\bigr)^2 - \mbox{$\frac{1}{4}$}\bigl(\Delta\Gamma_s^{}\bigr)^2
\,\,=\,\, 4\,\bigl|M^{12}_s\bigr|^2 - \bigl|\Gamma^{12}_s\bigr|^2 ~, \vphantom{\sum_i} \\
\Delta M_s^{}\,\Delta \Gamma_s^{} \,&=&\,
4\,\bigl| M^{12}_s\bigr|\,\bigl|\Gamma^{12}_s\bigr|\,\cos\phi_s^{} ~, \hspace{5ex}
\phi_s^{} \,\,=\,\, \arg\bigl(-M^{12}_s/\Gamma^{12}_s\bigr) ~, \label{DeltaGs} \\ &&
a_{\rm sl}^s \,\,=\,\, \frac{4\,\bigl|M^{12}_s\bigr|\,\bigl|\Gamma^{12}_s\bigr|\,\sin\phi_s^{}}
{4\,\bigl|M^{12}_s\bigr|^2 + \bigl|\Gamma^{12}_s\bigr|^2}   \label{assl}
\end{eqnarray}
in the notation of Ref.~\cite{Abazov:2010hv}.
The SM predicts~\cite{Lenz:2006hd,Kubo:2010mh}
\begin{eqnarray}
2\,M_s^{12,\rm SM} \,&=&\, 20.1 (1 \pm 0.40)\,e^{-0.035i}~{\rm ps}^{-1} ~, \hspace{5ex} \nonumber
2\,\bigl|\Gamma_s^{12,\rm SM}\bigr| \,\,=\,\, 0.096\pm 0.039~{\rm ps}^{-1}~, \\ \label{m12s_sm}
&& \phi_s^{\rm SM} \,\,=\,\, (4.2\pm 1.4)\times 10^{-3} \,\,=\,\, 0.24^\circ \pm 0.08^\circ ~.
\end{eqnarray}
Since  \,$\Delta\Gamma_s^{}\ll\Delta M_s^{}$\, and~\cite{Branco:1999fs}
\,$\bigl|\Gamma^{12}_s\bigr|\ll\bigl|M^{12}_s\bigr|$,
the commonly used expressions are
\begin{eqnarray}  \label{DeltaMs_approx}
\Delta M_s^{} \,\,\simeq\,\,  2\,\bigl|M^{12}_s\bigr| ~, \hspace{5ex}
\Delta\Gamma_s \,\,\simeq\,\,  2\,\bigl|\Gamma^{12}_s\bigr|\,\cos\phi_s^{} ~,
\end{eqnarray}
leading to
\begin{eqnarray} \label{assl_approx}
a_{\rm sl}^s \,\,\simeq\,\,
\frac{\bigl|\Gamma^{12}_s\bigr|\,\sin\phi_s^{}}{\bigl|M^{12}_s\bigr|}
\,\,\simeq\,\, \frac{2\,\bigl|\Gamma^{12}_s\bigr|\,\sin\phi_s^{}}{\Delta M_s^{}} ~.
\end{eqnarray}

The preceding equation for $a^s_{\rm sl}$ implies that any new physics which is to provide
a~successful explanation for the anomalous value of $a^s_{\rm sl}$ reported by D0 needs to
affect both $M_s^{12}$ and~$\Gamma_s^{12}$.
However, as Eqs.\,(\ref{DeltaMs_data}) and~(\ref{DeltaMs_approx}) indicate,
the magnitude of $M_s^{12}$ is strongly constrained by the experimental data, and
so the possible room for new physics lies mostly in $\Gamma_s^{12}$ and the relative
phase $\phi_s^{}$ between $M_s^{12}$ and $\Gamma_s^{12}$~\cite{Dobrescu:2010rh}.
A~related observation is that the smallness of the SM prediction $\phi_s^{\rm SM}$ suggests
that any new-physics effects which can significantly enhance $\phi_s^{}$ as well as
$\bigl|\Gamma_s^{12}\bigr|$ with respect to their SM values are likely to account for
the unexpectedly large value of~$a^{s,\rm exp}_{\rm sl}$.

Here we consider the possibility that this $a^s_{\rm sl}$ anomaly arises from
the contribution of a new particle of spin~one and mass under the $b$-quark mass.
Nonstandard spin-1 particles with masses of a few GeV or less have been explored to some
extent in various other contexts beyond the SM in the literature.
Their existence is in general still allowed by presently available data and also desirable,
as they may offer possible explanations for some of the recent experimental anomalies and
unexpected observations.
For instance, a spin-1 boson having mass of a few GeV and couplings to both quarks and
leptons has been proposed to explain the measured value of the muon $g$$-$2 and the NuTeV anomaly
simultaneously~\cite{Gninenko:2001hx}.
As another example, ${\cal O}$(MeV) spin-1 bosons which can interact with dark matter as well
as leptons may be responsible for the observed 511-keV emission from the Galactic bulge and
are potentially detectable by future neutrino telescopes~\cite{Hooper:2007jr}.
If its mass is of ${\cal O}$(GeV), such a particle may be associated with the unexpected excess
of positrons recently observed in cosmic rays, possibly caused by dark-matter
annihilation~\cite{Foot:1994vd}.
In the context of hyperon decay, a~spin-1 boson with mass around 0.2\,GeV,
flavor-changing couplings to quarks, and a dominant decay mode into $\mu^+\mu^-$
can explain the three anomalous events of \,$\Sigma^+\to p\mu^+\mu^-$\,
reported by the HyperCP experiment several years ago~\cite{He:2005we}.
Although in these few examples the spin-1 particles tend to have very small couplings to SM
particles, it is possible to test their presence in future high-precision
experiments~\cite{Hooper:2007jr,Foot:1994vd,He:2005we,Pospelov:2008zw}.
It~is therefore also interesting to explore a~light spin-1 boson as an explanation for
the $a^s_{\rm sl}$ anomaly.

In this paper we adopt a model-independent approach, assuming only that the spin-1
particle, which we shall refer to as~$X$, is lighter than the $b$ quark, carries no color
or electric charge, and has some simple form of flavor-changing interactions with quarks.
As we will elaborate, it is possible for $X$ with mass below the $b$-quark mass and
couplings satisfying current experimental constraints to yield a value of $a^s_{\rm sl}$
which is within the one-sigma range of the new D0 data.

\section{Interactions and amplitudes}

With $X$ being colorless and electrically neutral, we can express the Lagrangian describing
its effective flavor-changing couplings to $b$ and $s$~quarks as
\begin{eqnarray} \label{LbsX}
{\cal L}_{bsX}^{} \,\,=\,\,
-\bar s\gamma_\mu^{}\bigl(g_V^{}-g_A^{}\gamma_5^{}\bigr)b\,X^\mu \,\,+\,\, {\rm H.c.}
\,\,=\,\,
-\bar s\gamma_\mu^{}\bigl(g_{\rm L}^{}P_{\rm L}^{}+g_{\rm R}^{}P_{\rm R}^{}\bigr)b\,X^\mu
\,\,+\,\, {\rm H.c.} ~,
\end{eqnarray}
where $g_V^{}$ and $g_A^{}$ parametrize the vector and axial-vector couplings, respectively,
\,$g_{\rm L,R}^{}=g_V^{}\pm g_A^{}$,\, and \,$P_{\rm L,R}^{}=\frac{1}{2}(1\mp\gamma_5^{})$.\,
Generally, the constants $g_{V,A}^{}$ can be complex.
In principle, $X$~can have additional interactions, flavor-conserving and/or flavor-violating,
with other fermions which are parametrized by more coupling constants.
We assume that these additional parameters already satisfy other experimental constraints to
which they are subject, but with which we do not deal in this study.
Hence we will not consider much further phenomenological implications of such a~particle,
beyond those directly related to the D0 anomalous finding.
In the following, we derive the contributions of ${\cal L}_{bsX}^{}$ to
the amplitudes for several processes involving the $B_s$ meson.

For the mixing-matrix elements $M_s^{12}$ and $\Gamma_s^{12}$,
including the $X$ contributions we have
\begin{eqnarray} \label{M12_G12_X}
M_s^{12} \,\,=\,\, M_s^{12,\rm SM} + M_s^{12,X} ~, \hspace{5ex}
\Gamma_s^{12} \,\,=\,\, \Gamma_s^{12,\rm SM} + \Gamma_s^{12,X} ~.
\end{eqnarray}
To determine $M_s^{12,X}$, we apply the general relation
\,$2m_{B_s}^{}M_s^{12}=\bigl\langle B_s^0\bigr|{\cal H}_{b\bar s\to \bar b s}^{}
\bigl|\bar B_s^0\bigr\rangle$\,~\cite{Buchalla:1995vs}
to the effective Hamiltonian ${\cal H}_{b\bar s\to \bar b s}^X$ derived from the amplitude
for the tree-level transition \,$b\bar s\to\bar b s$\, mediated by $X$ in the $s$ and $t$
channels induced by~${\cal L}_{bsX}$.
Thus
\begin{eqnarray} \label{Hbs2bs}
{\cal H}_{b\bar s\to \bar b s}^X \,&=&\,
\frac{\bar s\gamma^\mu\bigl(g_{\rm L}^{}P_{\rm L}^{}+g_{\rm R}^{}P_{\rm R}^{}\bigr)b\,
      \bar s\gamma_\mu^{}\bigl(g_{\rm L}^{}P_{\rm L}^{}+g_{\rm R}^{}P_{\rm R}^{}\bigr)b}
{2\bigl(m_X^2-m_{B_s}^2\bigr)}
\nonumber \\ && +\,\,
\frac{\bigl\{\bar s\bigl[\bigl(g_{\rm L}^{}m_s^{}-g_{\rm R}^{}m_b^{}\bigr)P_{\rm L}^{} +
                  \bigl(g_{\rm R}^{}m_s^{}-g_{\rm L}^{}m_b^{}\bigr)P_{\rm R}^{}\bigr]b\bigr\}^2}
{2\bigl(m_X^2-m_{B_s}^2\bigr)m_X^2} ~,
\end{eqnarray}
where we have used in the denominators the approximation \,$p_X^2\simeq m_{B_s}^2\sim m_b^2$\,
appropriate for the $B_s$ rest-frame and included an overall factor of 1/2 to account for
the products of two identical operators.
This Hamiltonian was earlier obtained in a different context in Ref.~\cite{Oh:2009fm}.
In evaluating its matrix element at energy scales \,$\mu\sim m_b^{}$,\, one needs to include
the effect of QCD running from high energy scales which mixes different operators.
The resulting contribution of $X$ is
\begin{eqnarray} \label{M12sX}
M_s^{12,X} \,&=&\,
\frac{f_{B_s}^2\, m_{B_s}^{}}{3\bigl(m_X^2-m_{B_s}^2\bigr)} \Biggl[
\bigl(g_V^2+g_A^2\bigr) P_1^{\rm VLL} +
\frac{g_V^2\,\bigl(m_b^{}-m_s^{}\bigr)^2+g_A^2\,\bigl(m_b^{}+m_s^{}\bigr)^2}{m_X^2}\,
P_1^{\rm SLL}
\nonumber \\ && \hspace{16ex} +\,\, \bigl(g_V^2-g_A^2\bigr) P_1^{\rm LR}
+ \frac{g_V^2\,\bigl(m_b^{}-m_s^{}\bigr)^2-g_A^2\,\bigl(m_b^{}+m_s^{}\bigr)^2}{m_X^2}\,
P_2^{\rm LR} \Biggr] ~,
\end{eqnarray}
where \,$P_1^{\rm VLL}=\eta_1^{\rm VLL} B_1^{\rm VLL}$,\,
\,$P_1^{\rm SLL}=-\mbox{$\frac{5}{8}$}\, \eta_1^{\rm SLL} R_{B_s} B_1^{\rm SLL}$,\, and
\,$P_j^{\rm LR}=-\mbox{$\frac{1}{2}$}\, \eta_{1j}^{\rm LR} R_{B_s} B_1^{\rm LR}
                      + \mbox{$\frac{3}{4}$}\, \eta_{2j}^{\rm LR} R_{B_s} B_2^{\rm LR}$,\,
\,$j\,=\,1,2$\,~\cite{Buras:2001ra},
with the $\eta$'s denoting QCD-correction factors, the $B$'s being bag parameters defined by
the matrix elements
$\bigl\langle B_s^0\bigr|\bar s\gamma^\mu P_{\rm L}^{}b\,
\bar s\gamma_\mu^{}P_{\rm L}^{}b \bigl|\bar B_s^0\bigr\rangle =
\bigl\langle B_s^0\bigr|\bar s\gamma^\mu P_{\rm R}^{}b\,
\bar s\gamma_\mu^{}P_{\rm R}^{}b \bigl|\bar B_s^0\bigr\rangle =
\mbox{$\frac{2}{3}$} f_{B_s}^2 m_{B_s}^2 B_1^{\rm VLL}$,\,
\,$\bigl\langle B_s^0\bigr|\bar s P_{\rm L}^{}b\,\bar s P_{\rm L}^{}b\bigl|\bar B_s^0\bigr\rangle
= \bigl\langle B_s^0\bigr|\bar sP_{\rm R}^{}b\,\bar sP_{\rm R}^{}b\bigl|\bar B_s^0\bigr\rangle
= -\mbox{$\frac{5}{12}$} f_{B_s}^2 m_{B_s}^2 R_{B_s} B_1^{\rm SLL}$,\,
\,$\bigl\langle B_s^0\bigr|\bar s\gamma^\mu P_{\rm L}^{}b\,
\bar s\gamma_\mu^{}P_{\rm R}^{}b\bigl|\bar B_s^0\bigr\rangle =
-\mbox{$\frac{1}{3}$} f_{B_s}^2 m_{B_s}^2 R_{B_s} B_1^{\rm LR}$,\, and
\,$\bigl\langle B_s^0\bigr|\bar s P_{\rm L}^{}b\,\bar s P_{\rm R}^{}b\bigl|\bar B_s^0\bigr\rangle
=\mbox{$\frac{1}{2}$} f_{B_s}^2 m_{B_s}^2 R_{B_s} B_2^{\rm LR}$,\,
and \,$R_{B_s}=m_{B_s}^2/\bigl(m_b^{}+m_s^{}\bigr){}^2$.\,

As for $\Gamma_s^{12}$, it is in general affected by any physical state $f$ into which both
$B_s$ and $\bar B_s$ can decay.
Mathematically, $\Gamma_s^{12}$ is given by~\cite{Branco:1999fs}
\begin{eqnarray}
\Gamma_s^{12} \,\,=\,\,
\mbox{$\sum_f'$}\bigl({\cal M}(B_s\to f)\bigr)^*{\cal M}\bigl(\bar B_s\to f\bigr) ~,
\end{eqnarray}
the prime indicating that final-state kinematical factors and integrations are to be properly
incorporated.
In the SM, this is dominated by the CKM-favored \,$b\to c\bar c s$\, tree-level
processes~\cite{Lenz:2006hd}.
In contrast, with the $X$ mass \,$m_X^{}<m_b^{}$,\, the dominant processes contributing
to $\Gamma_s^{12,X}$ arise from decays induced by \,$b\,\bigl(\bar b\bigr)\to s\,(\bar s)\,X$,\,
such as \,$\bar B_s\,(B_s)\to\eta X$,\, \,$\bar B_s\,(B_s)\to\eta'X$,\,
and \,$\bar B_s\,(B_s)\to\phi X$.\,
It follows that $\Gamma_s^{12,X}$ can be written as
\begin{eqnarray} \label{G12sX}
\Gamma_s^{12,X} \,\,=\,\,
\mbox{$\sum_{f_X^{}}'$}\bigl({\cal M}(B_s\to f_X)\bigr)^*{\cal M}\bigl(\bar B_s\to f_X\bigr) ~,
\end{eqnarray}
where \,$f_X^{}=\eta X,\,\eta'X,\,\phi X,\ldots$\, for kinematically allowed~\,$B_s\to f_X^{}$.\,
Now, apart from the presence of squares of the coupling constants, $g_{V,A}^2$,
instead of their absolute values,
this sum is the same in form as the sum of rates \,$\Sigma_{f_X}\Gamma(B_s\to f_X)$,\,
which is approximately equivalent to the rate $\Gamma(b\to sX)$ of the inclusive
decay~\,$b\to s X$\, for \,$m_X^{}<m_b^{}-m_s^{}$.\,
Accordingly, one can rewrite $\Gamma_s^{12,X}$ using the formula for $\Gamma(b\to sX)$
derived from ${\cal L}_{bsX}$ above, with $|g_{V,A}^{}|^2$ replaced with $g_{V,A}^2$.
Thus
\begin{eqnarray} \label{G12sX_inc}
\Gamma_s^{12,X} \,&\simeq&\, \frac{\bigl|{\vec p}_X^{}\bigr|}{8\pi\,m_b^2 m_X^2}
\Bigl\{g_V^2\Bigl[\bigl(m_b^{}+m_s^{}\bigr)^2+2 m_X^2\Bigr]
\Bigl[\bigl(m_b^{}-m_s^{}\bigr)^2-m_X^2\Bigr] \nonumber \\ && \hspace*{12ex} +\,\,
g_A^2\Bigl[\bigl(m_b^{}-m_s^{}\bigr)^2+2 m_X^2\Bigr]
\Bigl[\bigl(m_b^{}+m_s^{}\bigr)^2-m_X^2\Bigr]\Bigr\} ~,
\end{eqnarray}
where ${\vec p}_X^{}$ is the 3-momentum of $X$ in the rest frame of~$b$.

As it turns out, however, for~\,$m_X^{}\mbox{\footnotesize$\,\gtrsim\,$}3$\,GeV\, we find that
$\Gamma_s^{12,X}$ evaluated using Eq.~(\ref{G12sX_inc}) is numerically less than that
using~Eq.~(\ref{G12sX}) with the sum being over \,$f_X=\eta X,\,\eta'X$, and~$\phi X$\,
alone.
This is an indication that the approximation in Eq.~(\ref{G12sX_inc}) is no longer good
for these larger values of $m_X^{}$, as soft QCD effects are no longer negligible in
relating \,$b\to sX$\, to the corresponding $\bar B_s$ process.
To get around this problem, for~\,$m_X^{}\mbox{\footnotesize$\,\gtrsim\,$}3$\,GeV\, we take
$\Gamma_s^{12,X}$ to be that given by Eq.~(\ref{G12sX}) with the sum being over
\,$f_X^{}=\eta X,\eta'X,\phi X$\, and neglect the effects of states $f_X^{}$
involving mesons heavier than the $\phi$ due to the smaller phase space of those states.
Hence, to evaluate $\Gamma_s^{12,X}$ in this case requires
the \,$\bar B_s\to(\eta,\eta',\phi)$\, matrix elements of the \,$b\to s$ operators
in~${\cal L}_{bsX}$,
which we expect take into account, at least partly,
the soft QCD effects not included in~Eq.~(\ref{G12sX_inc}).
The matrix element relevant to \,$\bar B_s \to P X$,\, with \,$P=\eta$ or $\eta'$,\, is
\begin{eqnarray}
\varepsilon_X^{*\mu}\bigl\langle P\bigl(p_P^{}\bigr)\bigr|\bar s\gamma_\mu^{} b
\bigl|\bar B_s \bigl(p_{B_s}^{}\bigr) \bigr\rangle \,\,=\,\,
2\varepsilon_X^*\cdot p_P^{}\, F_1^{B_s P} ~,
\end{eqnarray}
where \,$k=p_{B_s}^{}-p_P^{}=p_X^{}$,\, the form-factor $F_1^{B_s P}$ depends on~\,$k^2=m_X^2$,\,
and we have used the fact that the $X$ polarization $\varepsilon_X^{}$ and momentum $p_X^{}$
satisfy the relation \,$\varepsilon_X^*\cdot p_X^{}=0$.\,
For \,$\bar B_s \to\phi X$\,  we need
\begin{eqnarray}
\varepsilon_X^{*\mu}\bigl\langle \phi\bigl(p_\phi^{}\bigr)\bigr|\bar s\gamma_\mu^{}b
\bigl|\bar B_s \bigl(p_{B_s}^{}\bigr)\bigr\rangle \,&=&\,
\frac{2 V^{B_s \phi}}{m_{B_s}^{}+m_\phi^{}}\, \epsilon_{\mu\nu\sigma\tau}^{}\,
\varepsilon_X^{*\mu}\varepsilon_\phi^{*\nu} p_{B_s}^\sigma\, p_\phi^\tau ~, \\
\varepsilon_X^{*\mu}\bigl\langle \phi\bigl(p_\phi^{}\bigr)\bigr|\bar s\gamma_\mu^{}\gamma_5^{}b
\bigl|\bar B_s \bigl(p_{B_s}^{}\bigr)\bigr\rangle  \,&=&\,
i A_1^{B_s \phi}\, \bigl(m_{B_s}^{}+m_\phi^{}\bigr)\,\varepsilon_X^*\cdot\varepsilon_\phi^*
- \frac{2i A_2^{B_s \phi}\, \varepsilon_\phi^{*}\!\cdot\!k}{m_{B_s}^{}+m_\phi^{}}\,
\varepsilon_X^*\cdot p_\phi^{} ~,
\end{eqnarray}
where \,$k=p_{B_s}^{}-p_\phi^{}=p_X^{}$,\, and the form-factors $V^{B_s \phi}$ and
$A_{1,2}^{B_s \phi}$ are all functions of~\,$k^2=m_X^2$.\,
The amplitudes for \,$\bar B_s \to P X$\, and \,$\bar B_s\to\phi X$\, are then
\begin{eqnarray} \label{M_B2PX}
&& {\cal M}(\bar B_s \to P X) \,\,=\,\, 2\, g_V^{}\, F_1^{B_s P}\,
\varepsilon_X^*\!\cdot\!p_P^{} ~, \\ \label{M_B2VX}
{\cal M}(\bar B_s \to \phi X) \,&=&\,
-i g_A^{} \Biggl[
A_1^{B_s \phi}\, \bigl(m_{B_s}^{}+m_\phi^{}\bigr)\,\varepsilon_\phi^*\!\cdot\!\varepsilon_X^* \,-\,
\frac{2A_2^{B_s \phi}\, (\varepsilon_\phi^*\!\cdot\!p_X^{}) \, (\varepsilon_X^*\!\cdot\!p_\phi^{})}
{m_{B_s}^{}+m_\phi^{}} \Biggr]
\nonumber \\ && +\,\,
\frac{2 g_V^{}\, V^{B_s \phi}}{m_{B_s}^{}+m_\phi^{}}\, \epsilon_{\mu\nu\sigma\tau}^{}\,
\varepsilon_\phi^{*\mu}\varepsilon_X^{*\nu}p_\phi^\sigma\,p_X^\tau ~.
\end{eqnarray}
It follows that for~\,$m_X^{}\mbox{\footnotesize$\,\gtrsim\,$}3$\,GeV\, we have
\begin{eqnarray} \label{G12sX_exc}
\Gamma_s^{12,X} \,&\simeq&\,
\Gamma_s^{12,X}(\eta X)+\Gamma_s^{12,X}(\eta'X)+\Gamma_s^{12,X}(\phi X)~, \vphantom{\sum_i} \\
\Gamma_s^{12,X} (P X) \,&=&\, \frac{g_V^2\,\bigl|\vec p_P^{}\bigr|^3}{2\pi\, m_X^2}
\bigl(F_1^{B_s P}\bigr)^2 ~, \hspace{5ex}
\Gamma_s^{12,X} (\phi X) \,\,=\,\, \frac{\bigl|\vec p_{\phi}^{}\bigr|}{8\pi\, m_{B_s}^2}
\bigl( H_0^2 +H_+^2 +H_-^2 \bigr)
\end{eqnarray}
in the $B_s$ rest-frame, where~\cite{Kramer:1991xw}
\,$H_0^{}=-a x -b \bigl(x^2 -1\bigr)$\, and \,$H_{\pm}^{}=a \pm c \sqrt{x^2 -1}$,\, with
\begin{eqnarray}
a \,\,=\,\, g_A^{} A_1^{B_s \phi}\, \bigl(m_{B_s}^{} +m_\phi^{}\bigr) ~, && \hspace{3ex}
b \,\,=\,\, -\frac{2g_A^{}A_2^{B_s\phi}\,m_\phi^{}m_X^{}}{m_{B_s}^{}+m_\phi^{}} ~, \hspace{5ex}
c \,\,=\,\, -\frac{2 g_V^{}V^{B_s\phi}\,m_\phi^{} m_X^{}}{m_{B_s}^{} + m_\phi^{}} ~, ~~~~ \\
x \,\,=\,\, \frac{m_{B_s}^2 - m_\phi^2 - m_X^2}{2 m_\phi^{} m_X^{}} ~, && \hspace{3ex}
\bigl|\vec p_M^{}\bigr| \,\,=\,\, \frac{1}{2 m_{B_s}^{}}
\Bigl[ \bigl(m_{B_s}^2 +m_M^2 - m_X^2\bigr)^2 - 4 m_{B_s}^2 m_M^2 \Bigr]^{1/2} ~,
\end{eqnarray}
and $F_1^{B_s P}$, $A_{1,2}^{B_s \phi}$, $V^{B_s \phi}$ all evaluated at
\,$k^2=m_X^2$.\,
For numerical work in the next section, we employ
\,$F_1^{B_s\eta}\bigl(k^2\bigr)=-F_1^{B_dK}\bigl(k^2\bigr)\,\sin\varphi$,\, and
\,$F_1^{B_s\eta'}\bigl(k^2\bigr)=F_1^{B_dK}\bigl(k^2\bigr)\,\cos\varphi$\,~\cite{Carlucci:2009gr},
with \,$\varphi=39.3^\circ$\,~\cite{Feldmann:1998vh} and
the \,$B_d\to K$\, form-factor $F_1^{B_dK}\bigl(k^2\bigr)$ from Ref.~\cite{Cheng:2003sm},
as well as $A_{1,2}^{B_s\phi}\bigl(k^2\bigr)$ and $V^{B_s\phi}\bigl(k^2\bigr)$
from~Ref.~\cite{Ball:2004rg}.

\section{Numerical analysis\label{results}}

\begin{figure}[b] \vspace*{1ex}
\includegraphics[width=2.2in]{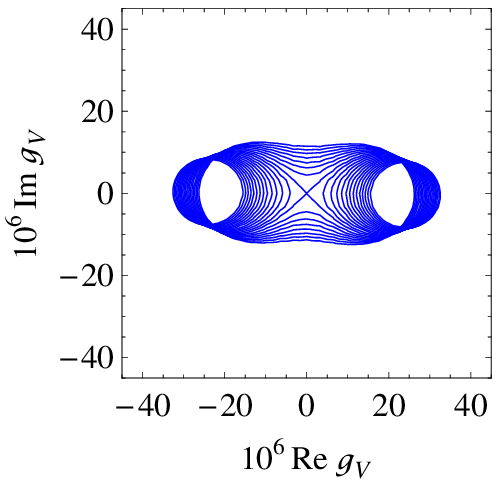} \, \, \,
\includegraphics[width=2.2in]{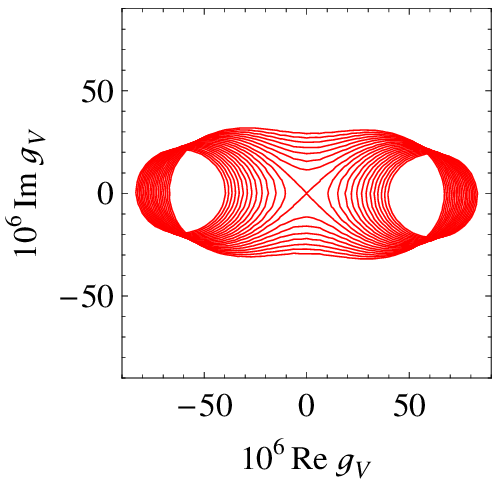}\vspace*{-1ex}
\caption{Regions of ${\rm Re}\,g_V^{}$ and ${\rm Im}\,g_V^{}$ allowed by
\,$\Delta M_s^{\rm exp}=2\bigl|M_s^{12}\bigr|$\, constraint
for \,$m_X^{}=2$\,GeV (left plot) and \,$m_X^{}=4$\,GeV (right plot)
under the assumption \,$g_A^{} = 0$.\label{ms12}} \vspace*{-3ex}
\end{figure}

We start with the constraints imposed by $\Delta M_s^{\rm exp}$ in~Eq.\,(\ref{DeltaMs_data}).
In this case, it is appropriate to use the approximate formula
\,$\Delta M_s^{}\simeq2\,\bigl|M^{12}_s\bigr|$,\, from Eq.\,(\ref{DeltaMs_approx}),
with $M_s^{12}$ given in Eq.\,(\ref{M12_G12_X}) and the $X$ contribution in~Eq.\,(\ref{M12sX}).
For numerical inputs, we adopt the SM numbers in Eq.\,(\ref{m12s_sm}),
\,$f_{B_s}=240$\,MeV,\, \,$m_b^{}\bigl(m_b^{}\bigr)=4.20$\,GeV,\,
\,$m_s^{}\bigl(m_b^{}\bigr)=80$\,MeV\,~\cite{Lenz:2006hd,Kubo:2010mh},
\,$P_1^{\rm VLL}=0.84$,\, \,$P_1^{\rm SLL}=-1.47$,\, \,$P_1^{\rm LR}=-1.62$,\,
\,$P_2^{\rm LR}=2.46$\,~\cite{Buras:2001ra},
and meson masses from Ref.~\cite{pdg}.
In Fig.\,\ref{ms12} we show the ranges of ${\rm Re}\,g_V^{}$ and ${\rm Im}\,g_V^{}$
satisfying \,$\Delta M_s^{\rm exp}=2\,\bigl|M_s^{12}\bigr|$\,
for \,$m_X^{}=2$ and 4~GeV, respectively, where for simplicity we have set
the coupling $g_A^{}$ to zero.
The contours in each of the plots correspond to variations of the SM contribution
$M_s^{12,\rm SM}$, which has an error of~40\%, as quoted in Eq.~(\ref{m12s_sm}).
Evidently, both ${\rm Re}\,g_V^{}$ and ${\rm Im}\,g_V^{}$ can be as large as
a few times~$10^{-5}$.
Assuming \,$g_V^{}=0$\, instead, we get allowed regions for ${\rm Re}\,g_A^{}$ and
${\rm Im}\,g_A^{}$ which are roughly almost three times smaller, with the vertical and
horizontal axes interchanged.
These restrictions from \,$\Delta M_s^{\rm exp}=2\bigl|M_s^{12}\bigr|$\, turn out to be
weaker than the ones we consider below using other $B_s$ observables.

Before proceeding, it is of interest also at this point to see how $g_{V,A}^{}$ may
compare to the analogous flavor-changing couplings $\bar g_{V,A}^{}$ of $X$ to
a pair of $d$ and $s$ quarks, subject to constraints from kaon-mixing data.
For definiteness, we take \,$m_X^{}=2$\,GeV,\, which is one of the values considered in our
numerical examples.
In this case, the pertinent observables are the mass difference  $\Delta M_K^{}$  between
$K_L^{}$ and $K_S^{}$ and the $CP$-violation parameter $\epsilon_K^{}$,
which are related to the mass matrix element \,$M_K^{12}=M_K^{12,\rm SM}+M_K^{12,X}$\, by
\,$\Delta M_K^{}=2\,{\rm Re}\,M_K^{12}+\Delta M_K^{\rm LD}$\,  and
\,$\epsilon_K^{}={\rm Im}\,M_K^{12}/\bigl(\sqrt2\,\Delta M_K^{\rm exp}\bigr)$,\,
where $M_K^{12,\rm SM}$ $\bigl(M_K^{12,X}\bigr)$ parameterizes the short-distance SM~($X$)
contribution and $\Delta M_K^{\rm LD}$ contains long-distance effects~\cite{Buchalla:1995vs}.
The SM can accommodate the measured value
\,$\Delta M_K^{\rm exp}=(3.483\pm0.006)\times10^{-12}$\,MeV\,~\cite{pdg},\, although
the calculation of $\Delta M_K^{\rm LD}$ suffers from significant
uncertainties~\cite{Buchalla:1995vs}, whereas the SM prediction
\,$|\epsilon_K^{}|_{\rm SM}^{}=\bigl(2.01^{+0.59}_{-0.66}\bigr)\times10^{-3}$\,~\cite{ckmfit}\,
agrees well with the data,
\,$|\epsilon_K^{}|_{\rm exp}^{}=(2.228\pm0.011)\times10^{-3}$\,~\cite{pdg}.
Accordingly, it is reasonable to require the $X$ contributions to satisfy
\,$2\,{\rm Re}M_K^{12,X}<3.4\times10^{-12}$\,MeV\, and
\,$\bigl|{\rm Im}\,M_K^{12,X}\bigr|/\bigl(\sqrt2\,\Delta M_K^{\rm exp}\bigr)<0.7\times10^{-3}$.\,
The expression for $M_K^{12,X}$ can be obtained from Eq.\,(\ref{M12sX}) after making
the appropriate replacements, namely with the new numbers
\,$f_K^{}=160$\,MeV,\, \,$m_K^{}=498$\,MeV,\, \,$m_s^{}(\mu)=115$\,MeV,\,
\,$m_d^{}/m_s^{}\simeq0$,\, \,$P_1^{\rm VLL}=0.48$,\, \,$P_1^{\rm SLL}=-18.1$,\,
\,$P_1^{\rm LR}=-36.1$,\, and \,$P_2^{\rm LR}=59.3$\,
at the scale \,$\mu=2$\,GeV\,~\cite{Buras:2001ra}.
With \,$m_X^{}=2$\,GeV\, and the above requirements on the $X$ contributions,
assuming \,$\bar g_A^{}=0$\, then leads one to
\,$\bigl({\rm Im}\,\bar g_V^{}\bigr){}^2-\bigl({\rm Re}\,\bar g_V^{}\bigr){}^2
\lesssim 4\times10^{-14}$\, and
\,$\bigl|\bigl({\rm Re}\,\bar g_V^{}\bigr)\bigl({\rm Im}\,\bar g_V^{}\bigr)\bigr|
\lesssim 4\times10^{-17}$,\,
implying that \,$\bigl|\bar g_V^{}\bigr|\lesssim 2\times10^{-7}$.\,
Similarly, setting \,$\bar g_V^{}=0$\, instead, one extracts
\,$\bigl|\bar g_A^{}\bigr|\lesssim 2\times10^{-7}$.\,
These bounds on $\bar g_{V,A}^{}$ from kaon data are much stronger than those on $g_{V,A}^{}$
derived from $\Delta M_s^{\rm exp}$ in the previous paragraph.
However, as we will see in the following, the corresponding values of $g_{V,A}^{}$ that can
reproduce the D0 measurement are much smaller and have bounds roughly similar to
these $\bar g_{V,A}^{}$ numbers.
Thus, although in our model-independent approach $\bar g_{V,A}^{}$ are not necessarily
related to $g_{V,A}^{}$, this exercise serves to illustrate that in models where the two
sets of couplings are expected to be comparable in size it is possible
to satisfy both kaon-mixing and $B_s$ data.

To explore the ranges of $g_{V,A}^{}$ allowed by the other $B_s$ quantities,
$\Delta\Gamma_s^{\rm exp}$ and $a_{\rm sl}^{s,\rm exp}$, besides~$\Delta M_s^{\rm exp}$,
their values being quoted in Eqs.\,(\ref{asslx}) and~(\ref{DeltaMs_data}),
we employ the exact relations written in Eqs.\,(\ref{DeltaGs}) and~(\ref{assl}),
although one would arrive at similar results with the approximate formulas in
Eqs.\,(\ref{DeltaMs_approx}) and~(\ref{assl_approx}).
The relevant SM numbers are listed in Eq.~(\ref{m12s_sm}).
For the $X$ contributions, we have $M_s^{12,X}$ in Eq.~(\ref{M12sX}), whereas
$\Gamma_s^{12,X}$ is from Eq.~(\ref{G12sX_inc}) if \,$m_X^{}<3$\,GeV\, and
from Eq.~(\ref{G12sX_exc}) otherwise.
To simplify our analysis, we again assume only one of $g_{V,A}^{}$ to
be contributing at a time, setting the other one to zero.

We also need to take into account the inclusive decay \,$b\to s X$\, because
it provides constraints on $g_{V,A}^{}$ via its contribution, $\Gamma(b\to sX)$,
to the $B_s$ total-width~$\Gamma_{B_s}$.
Though the experimental value of $\Gamma_{B_s}$ is fairly well determined,
\,$\Gamma_{B_s}^{\rm exp}=0.70\pm0.02{\rm\,ps}^{-1}$\,~\cite{pdg},
its theoretical prediction in the SM involves significant uncertainties,
mainly due to $\Gamma_{B_s}^{\rm SM}$ being proportional to $m_b^5$ at leading order in
the $1/m_b^{}$ expansion~\cite{Bigi:1992su}.
With $m_b^{}$ having its PDG value~\cite{pdg}, the error of $\Gamma_{B_s}^{\rm SM}$
from $m_b^5$ alone would be of order~20\%.
There are additional uncertainties from the $f_{B_s}^2$ dependence of
\,$\Gamma_{B_s}^{\rm SM}$,\, as \,$f_{B_s}=240\pm40$\,MeV\,~\cite{Kubo:2010mh},
but they occur at subleading order in the  $1/m_b^{}$  expansion~\cite{Gabbiani:2004tp}.
Conservatively, we then require $\Gamma(b\to sX)$ to be smaller than
\,$0.15\,\Gamma_{B_s}\simeq0.1{\rm\,ps}^{-1}$,\,
but we will also assume, alternatively, the somewhat bigger upper-bound
of~\,$0.15{\rm\,ps}^{-1}$.\,
We will comment on the implications of $\Gamma(b\to sX)$ bounds stricter than
\,$\Gamma(b\to sX)<0.1{\rm\,ps}^{-1}$\, as well.

We remark that the same $\Gamma(b\to sX)$ also contributes to the total widths $\Gamma_{B_d}$
and $\Gamma_{B_u}$ of the $B_d$ and $B_u^+$~mesons, respectively, the SM calculations of which
involve sizable uncertainties similar to that of~$\Gamma_{B_s}^{\rm SM}$.
Since the SM predicts the width ratios \,$\Gamma_{B_d}/\Gamma_{B_s}$\, and
\,$\Gamma_{B_d}/\Gamma_{B_u}$\, to be only a few percent away from unity~\cite{Gabbiani:2004tp},
it follows that the $\Gamma(b\to sX)$ contributions to $\Gamma_{B_s,B_d,B_u}$ respect
the experimental numbers \,$\Gamma_{B_d}/\Gamma_{B_s}=1.05\pm0.06$\, and
\,$\Gamma_{B_d}/\Gamma_{B_u}=1.071\pm0.009$\,~\cite{pdg}.

\begin{figure}[b]
\includegraphics[width=174mm]{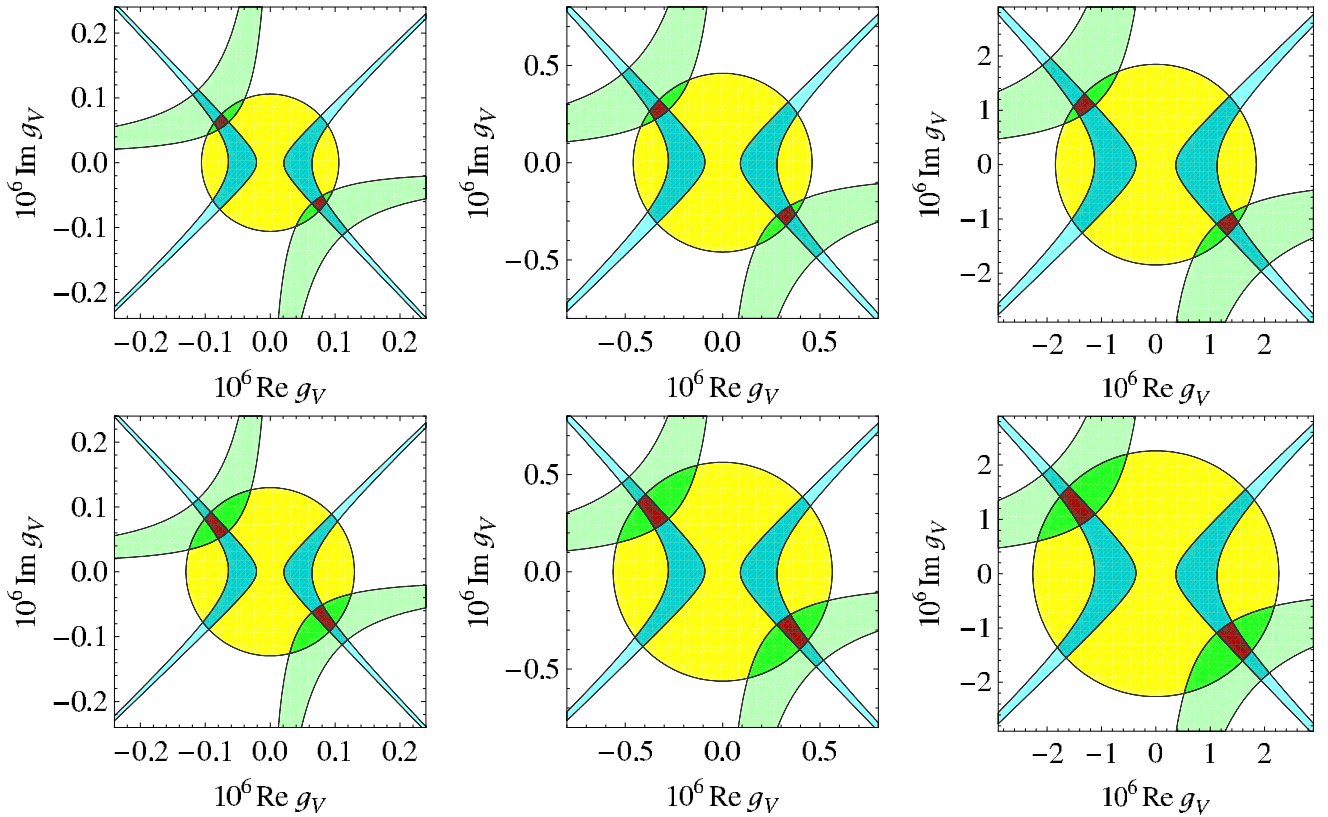}
\caption{Regions of ${\rm Re}\,g_V^{}$ and ${\rm Im}\,g_V^{}$ allowed by
$a_{\rm sl}^{s,\rm exp}$ constraint (green), $\Delta M_s^{\rm exp}\Delta\Gamma_s^{\rm exp}$
constraint (blue), \,$\Gamma(b\to s X)<0.1{\rm\,ps}^{-1}$ (yellow), and all of them (dark red) for
\,$m_X^{}=0.5$\,GeV~(upper left plot), 2\,GeV~(upper middle plot), and 4\,GeV~(upper right plot),
under the assumption \,$g_A^{}=0$.
The lower plots are the same as the upper ones, except that
\,$\Gamma(b\to s X)<0.15{\rm\,ps}^{-1}$.\label{allowed}}
\end{figure}

In Fig.~\ref{allowed}, we display the allowed values of ${\rm Re}\,g_V^{}$ and
${\rm Im}\,g_V^{}$, assuming  $g_A^{}=0$,\, subject to the requirements from
the one-sigma ranges of $\Delta M_s^{\rm exp}$, $\Delta\Gamma_s^{\rm exp}$, and
$a_{\rm sl}^{s,\rm exp}$ applied in Eqs.\,(\ref{DeltaGs}) and~(\ref{assl}),
plus the restrictions on~$\Gamma(b\to s X)$.
For the reasons described in the preceding section, in drawing this figure
we have employed the expression for $\Gamma(b\to sX)$ from $\Gamma_s^{12,X}$ in
Eq.~(\ref{G12sX_inc}) if \,$m_X^{}<3$\,GeV\, and Eq.~(\ref{G12sX_exc}) otherwise,
with $g_V^2$ and $g_A^2$ replaced by $|g_V^{}|^2$ and $|g_A^{}|^2$, respectively.
Choosing \,$m_X^{}=0.5,\,2,\,4$~GeV\, for illustration, we have
imposed~\,$\Gamma(b\to s X)<0.1{\rm\,ps}^{-1}$\, in the upper plots
and \,$\Gamma(b\to s X)<0.15{\rm\,ps}^{-1}$\, in the lower ones.
On each plot, the (blue) regions satisfying the $\Delta M_s^{\rm exp}\Delta\Gamma_s^{\rm exp}$
constraint lie on all the four quadrants and are narrower than the (green) regions
satisfying $a_{\rm sl}^{s,\rm exp}$, which lie on only the second and fourth quadrants.
The circular (yellow) regions represent the $\Gamma(b\to s X)$ bounds.
Clearly, there is parameter space of $X$ (in dark red) that can cover part of the one-sigma
range of $a_{\rm sl}^{s,\rm exp}$ and is simultaneously allowed by the other two constraints.
In each $m_X^{}$ case, the overlap area allowed by all the constraints is significantly
larger with the less restrictive bound~\,$\Gamma(b\to s X)<0.15{\rm\,ps}^{-1}$.\,
Evidently, the size of each of the areas corresponding to the different constraints
is sensitive to the value of $m_X^{}$ and increases as the latter grows.
The overlap region satisfying all the constraints also increases in size
with~$m_X^{}$.

\begin{figure}[t] %\vspace*{2ex}
\includegraphics[width=2.3in]{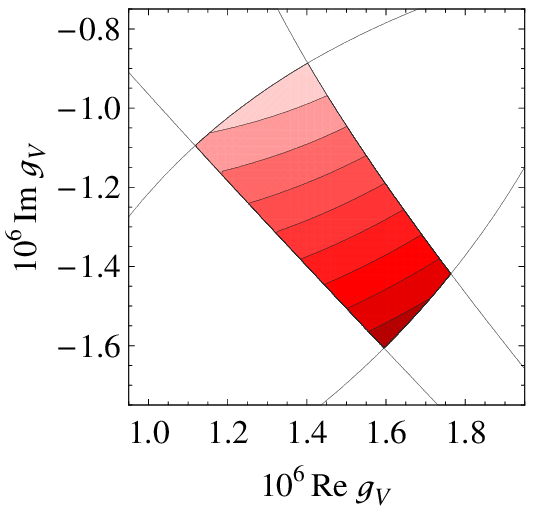} \,
\includegraphics[width=2.3in]{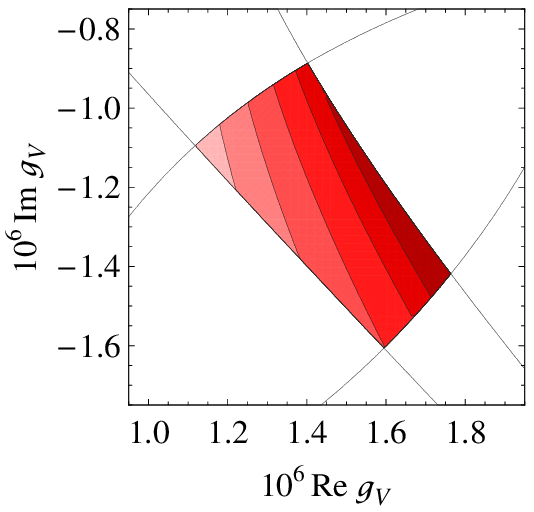} \vspace*{-1ex}
\caption{Values of $|\Gamma_s^{12}|$ (left plot) and $\sin\phi_s^{}$
(right plot) for \,$m_X^{}=4$\,GeV\, and the $\bigl({\rm Re}\,g_V^{},{\rm Im}\,g_V^{}\bigr)$
overlap region in the fourth quadrant of the lower-right plot in Fig.~\ref{allowed}
allowed by all the constraints, with \,$\Gamma(b\to s X)<0.15{\rm\,ps}^{-1}$.\,
In the left plot, from darkest to lightest, the differently shaded (red colored) areas
correspond to
\,$\bigl|\Gamma_s^{12}/\Gamma_{s}^{12,\rm SM} \bigr|>3.1,\,2.9,\,2.7,\ldots,\,1.5$,\,
respectively, with each region including the area of the next darker region
and $\bigl|\Gamma_{s}^{12,\rm SM}\bigr|$ being its central value.
Similarly, in the right plot, from darkest to lightest
\,$\sin\phi_s^{}< -0.99,\,-0.98,\,-0.96,\,-0.93,\,-0.89,\,-0.85$.\,\label{values}}
\end{figure}

To illustrate in more detail the impact of $X$ on the values of $|\Gamma_s^{12}|$ and
$\sin\phi_s^{}$ corresponding to the parameter space allowed by all the constraints,
we display the graphs in Fig.~\ref{values} in the case of \,$m_X^{}=4$\,GeV\,
and~\,$\Gamma(b\to s X)<0.15{\rm\,ps}^{-1}$.\,
These (red) shaded regions are none other than the (dark red) overlap region in the fourth
quadrant of the lower-right plot in Fig.~\ref{allowed}, but
one could alternatively use the overlap region in the second quadrant.
The left plot in Fig.~\ref{values} indicates that the size of $\bigl|\Gamma_s^{12}\bigr|$
can be enhanced to \,3.1\, times the central value of~$|\Gamma_{s}^{12,\rm SM}|$.
Furthermore, from the right plot, the magnitude of $\sin\phi_s^{}$ can be increased to almost~1,
which is roughly a few hundred times larger than its SM value.
Combining them leads to
\,$-0.016\,\mbox{\footnotesize$\lesssim$}\,a^s_{\rm sl}\,
\mbox{\footnotesize$\lesssim$}\,$$-$0.007.\,
Thus the enhancement of \,$|\Gamma_s^{12}|\sin\phi_s^{}$\, generated by the $X$ contribution
can be sufficiently sizable to yield a prediction for $a^s_{\rm sl}$ which can reach most of
the one-sigma range of the anomalous $a^{s,\rm exp}_{\rm sl}$, including its central value.
For lower values of $m_X^{}$, the situations are similar, as can be inferred from the lower
plots in Fig.~\ref{allowed}, although the allowed
$\bigl({\rm Re}\,g_V^{},{\rm Im}\,g_V^{}\bigr)$ areas are smaller.
With the more restrictive bound~\,$\Gamma(b\to s X)<0.1{\rm\,ps}^{-1}$,\, part of
the one-sigma range of $a^{s,\rm exp}_{\rm sl}$ can still be reproduced, as the upper
plots in Fig.~\ref{allowed} imply, but its central value is no longer reachable.

If we require a bound stricter than~\,$\Gamma(b\to s X)<0.1{\rm\,ps}^{-1}$,\,
then the $X$ contribution may not be able to lead to any of the one-sigma values
of~$a^{s,\rm exp}_{\rm sl}$.
However, in that case there can still be  $\bigl({\rm Re}\,g_V^{},{\rm Im}\,g_V^{}\bigr)$
regions allowed by all the constraints if one considers instead 90\%-C.L. ranges of
$a^{s,\rm exp}_{\rm sl}$, $\Delta M_s^{\rm exp}$, and~$\Delta\Gamma_s^{\rm exp}$.
More definite statements about this would have to await more precise data on $a^s_{\rm sl}$
from future experiments.

\begin{figure}[t]
\includegraphics[width=175mm]{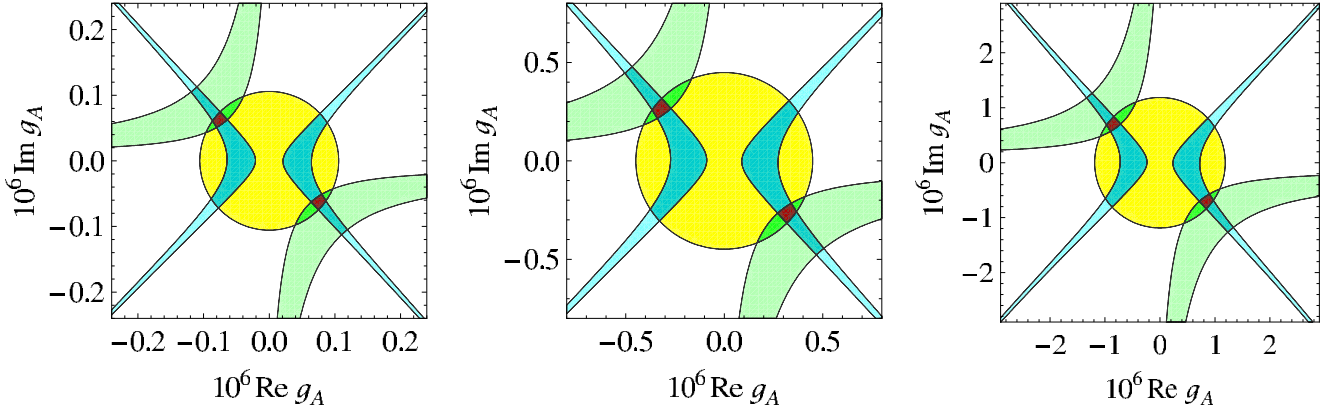}
\caption{Regions ${\rm Re}\,g_A^{}$ and ${\rm Im}\,g_A^{}$ allowed by
$a_{\rm sl}^{s,\rm exp}$ constraint (green), $\Delta M_s^{\rm exp}\Delta\Gamma_s^{\rm exp}$
constraint (blue), \,$\Gamma(b\to s X)<0.1{\rm\,ps}^{-1}$ (yellow), and all of them (dark red) for
\,$m_X^{}=0.5$\,GeV~(left plot), 2\,GeV~(middle plot), and 4\,GeV~(right plot),
under the assumption \,$g_V^{}=0$.\label{allowed_ga}}
\end{figure}

In the case that \,$g_V^{}=0$,\, we show in Fig.~\ref{allowed_ga} the values of
${\rm Re}\,g_A^{}$ and ${\rm Im}\,g_A^{}$ allowed by the constraints from
$a_{\rm sl}^{s,\rm exp}$ and $\Delta M_s^{\rm exp}\Delta\Gamma_s^{\rm exp}$
at the one-sigma level.
We have chosen \,$m_X^{}=0.5,\,2,\,4$~GeV\, as before, but imposed
only~\,$\Gamma(b\to s X)<0.1{\rm\,ps}^{-1}$.\,
The effects of $X$ here can be seen to be qualitatively similar to those in
the $g_V^{}\neq0$\, and $g_A^{}=0$\, case.

Finally, a few comments on distinguishing the scenario that we have proposed to reproduce
the D0 result from the other proposals in the literature seem to be in order.
Since the main feature in our proposal is the presence of a new light spin-1 boson with
flavor-changing couplings to $b$ and $s$ quarks, if the D0 finding is confirmed by other
experiments, the results of our model-independent study can serve to help motivate
experimentalists to look for the particle in various \,$b\to s$\, transitions,
such as by scrutinizing the dilepton-mass distributions in
\,$\bar B_s\to\eta^{(\prime)}\ell^+\ell^-,\,\phi\ell^+ \ell^-$,\, and
\,$\bar B_d\to\bar K^{(*)0}\ell^+\ell^-$\, in case it has sufficient branching
ratios into $\ell^+\ell^-$.
Since its couplings tend to be very small, the searches for the particle would require
a high degree of precision, which could hopefully be realized at LHCb or future $B$ factories.
If a new spin-1 particle is discovered in a~measurement of some \,$b\to s$\, transition,
to proceed and examine if the particle is the one that can reproduce the D0 anomaly,
it would be necessary to invoke model dependence, as different models containing such
a particle would likely have different values of the additional flavor-conserving and
flavor-violating couplings which the particle might have to various fermions,
subject to other experimental data.
The adoption of model specifics would also be unavoidable in order to distinguish this scenario
from other new-physics scenarios which could account for the D0 anomaly without
a~nonstandard light spin-1 boson, especially if it turned out to be experimentally elusive.
If a new light spin-1 particle were to be detected first outside the $B$ sector,
it would again be necessary to have a model to make connections to the $B$ sector.

\section{Conclusions\label{concl}}

We have investigated the possibility that the anomalous like-sign dimuon charge asymmetry
in semileptonic $b$-hadron decays recently measured by the D0 Collaboration arises from the
contribution of a light spin-1 particle, $X$, to the mixing of $B_s$ mesons.
Taking a model-independent approach, we have assumed only that $X$ is lighter than the $b$ quark,
carries no color or electric charge, and has vector and axial-vector $bsX$ couplings.
Thus, in contrast to a heavy $Z^{\prime}$ particle, $X$ can be produced as a physical particle
in $B_s$ decay, and so it affects not only the mass matrix element~$M_s^{12}$,
but also the decay matrix element~$\Gamma_s^{12}$.
We have found that the $X$ contribution can enhance the magnitude of $\Gamma_s^{12}$ as well as
the relative $CP$-violating phase $\phi_s^{}$ between $M_s^{12}$ and $\Gamma_s^{12}$
by a significant amount.
More precisely, taking into account experimental constraints from a~number of $B_s$ observables,
namely $\Delta M_s^{}$, $\Delta\Gamma_s^{}$, and $\Gamma_{B_s}$,
we have shown that the effect of $X$ can increase $|\Gamma_s^{12}|$ to become a few times greater
than its SM prediction and enlarge the size of $\sin\phi_s^{}$ by a factor of a few hundred.
As a consequence, the $X$ contribution can lead to a prediction for $a_{\rm sl}^s$ which is
consistent with its anomalous value as measured by D0 within one standard-deviation and
possibly even reaches its central value.
We have therefore demonstrated that a light spin-1 particle can offer a viable
explanation for the D0 anomaly.
Whether or not future $B_s$ experiments confirm the new D0 finding, the coming data will
likely be useful for further probing new-physics scenarios involving light spin-1 particles.

\acknowledgments

This work was supported in part by NSC and NCTS.
We thank X.G.~He and Alexander Lenz for helpful comments.

\end{document}